\documentclass[twocolumn, switch]{article} % Method A for two-column formatting

\usepackage{preprint}

\usepackage[numbers,square]{natbib}
\usepackage[utf8]{inputenc}	% allow utf-8 input
\usepackage[T1]{fontenc}	% use 8-bit T1 fonts
\usepackage{xcolor}		% colors for hyperlinks
\usepackage[colorlinks = true,
            linkcolor = purple,
            urlcolor  = blue,
            citecolor = cyan,
            anchorcolor = black,
            hypertexnames=false]{hyperref}	% Color links to references, figures, etc.
\usepackage{booktabs} 		% professional-quality tables
\usepackage{nicefrac}		% compact symbols for 1/2, etc.
\usepackage{microtype}		% microtypography
\usepackage{lineno}		% Line numbers
\usepackage{float}			% Allows for figures within multicol
\usepackage{amsmath, amsthm, amssymb, amsfonts}

\usepackage{newfloat}
\DeclareFloatingEnvironment[name={Supplementary Figure}]{suppfigure}
\usepackage{sidecap}
\sidecaptionvpos{figure}{c}

\usepackage{titlesec}
\titlespacing\section{0pt}{12pt plus 3pt minus 3pt}{1pt plus 1pt minus 1pt}
\titlespacing\subsection{0pt}{10pt plus 3pt minus 3pt}{1pt plus 1pt minus 1pt}
\titlespacing\subsubsection{0pt}{8pt plus 3pt minus 3pt}{1pt plus 1pt minus 1pt}

\usepackage{enumitem}
\setlist{leftmargin=1em}
\title{HOList: An Environment for Machine Learning of Higher-Order Theorem Proving}

\usepackage{authblk}

\author{Kshitij Bansal$^*$}
\author{Sarah M. Loos$^*$}
\author{Markus N. Rabe$^*$}
\author{Christian Szegedy$^*$}
\author{Stewart Wilcox$^*$}
\affil{\tt{$\{$kbk,smloos,mrabe,szegedy,stewbasic$\}$@google.com}\\Google Research, Mountain View, California, USA}

\begin{document}

\twocolumn[ % Method A for two-column formatting
  \begin{@twocolumnfalse} % Method A for two-column formatting

\maketitle

\begin{abstract}
We present an environment, benchmark, and deep learning driven automated theorem prover for higher-order logic.
Higher-order interactive theorem provers enable the formalization of arbitrary mathematical theories and thereby present an interesting, open-ended challenge for deep learning.
We provide an open-source framework based on the HOL Light theorem prover that can be used as a reinforcement learning environment.
HOL Light comes with a broad coverage of basic mathematical theorems on calculus and the formal proof of the Kepler conjecture, from which we derive a challenging benchmark for automated reasoning.
We also present a deep reinforcement learning driven automated theorem prover, DeepHOL, with strong initial results on this benchmark.
\end{abstract}
%\keywords{First keyword \and Second keyword \and More} % (optional)
\vspace{0.35cm}

\end{@twocolumnfalse} % Method A for two-column formatting
]

\section{Introduction}
\label{introduction}

Formalization of mathematics and the automated creation of new mathematical content is at the frontier of current AI techniques.
Given the fundamental nature of mathematics and its importance for most scientific disciplines, the capability for high level formal mathematical reasoning is both an important practical task as well as one of the most challenging case studies in AI.
However, traditional formal computer mathematics has been a fragmented domain, exploring various approaches for different logical foundations.
This has led to a large number of incompatible theorem proving systems, which added extra challenges for AI researchers trying to push the limits of formal reasoning using machine learning.

Well-defined, large-scale benchmarks were instrumental for unifying disparate efforts in machine learning research:
LibriSpeech~\cite{panayotov2015librispeech} for speech recognition,
the Netflix prize~\cite{bennett2007netflix} for recommendation,
ImageNet~\cite{deng2009imagenet} for object recognition,
MSCOCO~\cite{lin2014microsoft} for object detection and segmentation,
WMT~\cite{bojar2014findings} for machine translation, and
SQuAD~\cite{rajpurkar2016squad} for question answering - just to name a couple of examples.
Benchmarks have fostered collaboration and competition and provide a means to measure progress,
contributing significantly to accelerated progress and reproducible science.

This paper provides a benchmark and reinforcement learning environment for theorem proving.
The long-term goal is to enable the automatic formalization of large theories,
and hence we want to start with a theorem proving system that has a track-record of large-scale formalization efforts and includes a large corpus of foundational mathematics for benchmarking and learning.
Our choice fell on HOL Light, the interactive theorem prover (ITP) in which the proof of the Kepler conjecture~\cite{hales2017formal} has been formalized.
The formalization of the proof of the Kepler conjecture has been a huge effort, taking over 20 person-years to complete, and required formalizing a significant part of arithmetic, linear algebra, and multivariate analysis.
The resulting benchmark consists of 2199 definitions and 29462 theorems and lemmata, which capture a variety of interesting mathematics and should be a practical seed for new (auto-)formalization efforts.

To demonstrate the feasibility of the proposed learning task, we present an automated theorem prover powered by deep learning, called DeepHOL.
Based on a simple solver architecture, DeepHOL learns to prove theorems based on imitating human proofs and improves itself using reinforcement learning.
Given a proof goal (represented as a string) DeepHOL learns to predict the tactic (and its arguments) that leads to a successful proof.
Thereby, DeepHOL achieves theorem proving capabilities that are comparable to much more complicated state-of-the-art automated theorem proving systems.
In our open-source release, available at \url{http://deephol.org}, we expose the APIs of our modular theorem prover.
This simplifies the development of new provers significantly and allows researchers to focus on the machine learning aspects.

The contributions of our work are the following:
\begin{itemize}
\item An instrumented, pre-packaged version of HOL Light that can be used as a
  reinforcement learning environment for theorem proving using our well-defined,
  stable Python API. Our solution comes with optimized startup capabilities for
  proof search, while allowing replay and strict verification of the produced
  proofs.
\item Proof export and import capabilities that allow for managing large theories
  programmatically from the Python interface.
\item A full-fledged, competitive automated neural theorem proving system
   that can automatize theorem proving in higher-order logic at tactic level directly.
\item A large scale reinforcement learning system that was used for
  training our prover.
\item Comparison of neural model architectures for theorem proving purposes.
\item Well-defined benchmarks on our HOL Light based environment to enable
  research and measuring progress of AI driven theorem proving in large theories.
\end{itemize}

This paper is organized as follows.
We discuss related work in Section~\ref{sec:related} before we describe our theorem proving environment in Section~\ref{sec:ArchitectureOfTheEnvironment}.
In Section~\ref{sec:benchmark} we present the organization of the benchmark.
The DeepHOL automated theorem prover is described in Section~\ref{sec:deephol} and we discuss first experimental results for it in Section~\ref{sec:comparisons}. Then we conclude in Section~\ref{sec:conclusion}.

\section{Related Work}
\label{sec:related}

The earliest work of applying machine learning on reasoning in large theories is~\cite{urban2008malarea}.
The most most similar works to ours are TacticToe~\cite{gauthier2017tactictoe} and GamePad~\cite{huang2018gamepad}.
TacticToe is the first published result on machine learning tackling higher-order theorem proving at a relatively large scale at tactic level~\cite{gauthier2017tactictoe}.
Although TacticToe is a great success that came with significant improvements over previous automated theorem proving systems, they do not propose an easy to use benchmark or environment for machine learning researchers.
TacticToe does not employ deep learning nor reinforcement learning.
They rely on the HOL4~\cite{slind2008brief} system that has a significantly less theorems with more complex human proof scripts with a larger number of more elementary tactics.

GamePad has very similar objectives to ours~\cite{huang2018gamepad}.
They also provide an easy-to-use Python API for an interactive theorem prover, and they present test and training sets.
They chose to base their system on Coq~\cite{bertot2013interactive}, an interactive theorem prover based on the calculus of inductive constructions.
While enabling automatic code extraction, it comes with a much smaller coverage of fundamental mathematics.
Even including the formalization of the Feit-Thompson theorem, their benchmark comprises only 1602 theorems and lemmas, while ours features 29462 theorems and lemmas.
Besides presenting a much larger data set, we also demonstrate the feasibility of achieving state-of-the-art prover performance based on our data and environment by presenting a deep learning based theorem prover.
We also report the results as theorem proving performance instead of proxy metrics.

Other interactive theorem provers we could have based a learning environment on include Mizar~\cite{mml-homepage}, Isabelle~\cite{wenzel08isabelle}, HOL4~\cite{slind2008brief}, and Lean~\cite{de2015lean}.
The Mizar mathematical library is probably the most comprehensive formalization effort, but its declarative style makes it hard to employ proof search, and its source code is not freely available.
Like Coq and HOL Light, also Isabelle~\cite{wenzel08isabelle} was used for major formalization efforts, such as the formalization of the seL4 microkernel~\cite{Klein2009seL4}.
We are not aware of a comprehensive coverage of fundamental mathematics in Isabelle, HOL4, or Lean.

% One of the scientific uses of formalized mathematics is formal proof checking
% of highly complicated proofs.
% These can be extremely hard, if not impossible, for human peers to check fully.
% The most impressive examples are the machine checked proofs of the
% Feit--Thompson theorem~\cite{gonthier2013odd},
% the four color theorem~\cite{gonthier2008formal} and
% the Kepler conjecture~\cite{hales2017formal}.
% All three formalization efforts relied on interactive theorem provers (ITPs):
% Coq~\cite{Coq} in the case of Feit-Thompson and the four color theorem, and
% HOL Light~\cite{Harrison96} for the Kepler conjecture.
% These formalization efforts are huge projects.
% For example, the formalization of the proof of the Kepler conjecture took over $20$ person-years to complete, and required formalizing a significant part of arithmetic, linear algebra, and multivariate analysis.

In closely related work, \citet{kaliszyk2014learning} translate from HOL Light
and Flyspeck to automated theorem provers and SMT solvers, for which they learn a premise selector.
In contrast to our work, they use neither deep learning nor reinforcement learning.
Similar methods for premise selection on the HOL Light corpora were proposed in~\cite{kaliszyk2012initial}.

The first use of deep neural networks for large scale theorem proving was proposed in~\cite{alemi2016deepmath}.
They have used convolutional networks for premise selection in large theories, particularly on Mizar mathematical library~\cite{mml-homepage}.
Those methods were used as a pre-selection for applying the first order logic automated theorem prover E~\cite{Sch02-AICOMM}.
We have reused several ideas from that paper, including some aspects of our neural network architecture and the hard negative mining methodology.

\citet{whalen2016holophrasm} proposed a purely deep reinforcement learning based solution for theorem proving for the Metamath prover~\cite{megill1997metamath}.
This work was moderately successful, finding mostly proofs for very simple theorems, especially in propositional logic.
On the other hand, Metamath is not considered to be a serious contender for large scale mathematical formalization work.

\citet{loos2017deep} proposed deep neural networks to augment theorem prover E~\cite{Sch02-AICOMM} to rank given clauses during proof search.
Here, we propose a neural prover written from scratch, relying solely on a small set of preexisting tactics and neural networks for all high level decisions.

\citet{kaliszyk2017holstep} proposed a machine learning benchmark for higher-order logic reasoning based on the HOL Light corpus. It features a few static datasets and it remains unclear how performance of machine learning models on this dataset relates to real world prover performance.~\cite{kaliszyk2018reinforcement} demonstrated the viability of reinforcement learning with XGBoost and LIBLINEAR~\cite{fan2008liblinear} on hand engineered features in first order logic context using leanCoP~\cite{otten2003leancop} on Mizar mathematical library~\cite{mml-homepage}.

Earlier works on employing (non-deep) machine learning for theorem proving in general and for reasoning in large theories include~\cite{Sch00,duncan2004use,urban2011malecop,kuhlwein2012overview,kaliszyk2013stronger,kuhlwein2013mash,alama2014premise,BridgeHP14,kaliszyk2014machinelearner,kaliszyk2014machine,farber2015random,kaliszyk2015mizar,kaliszyk2015efficient,kaliszyk2015femalecop,kaliszyk2015learning,gauthier2015premise,blanchette2016learning}.
Recently,~\citet{wang2017premise} proposed a premise selection method utilizing deep graph embeddings.

\section{Architecture of the Environment}
\label{sec:ArchitectureOfTheEnvironment}

Here we describe the architecture of the evaluation and training environment.
The goal of the environment is to enable artificial agents to interact with the HOL Light interactive theorem prover (ITP) in a replicable manner.

\subsection{ITP Terminology}
In order to describe our changes to HOL Light, it is helpful to establish some common terminology.
To prove a theorem in an ITP, the human user starts with entering the theorem's statement as the \emph{goal} of a new proof.
The ITP provides a small number of \emph{tactics} to manipulate the goal.
Tactics may have \emph{tactic arguments}, which can be a previously proven theorem or a list of previously proven theorems.
(There are also tactics that take terms as arguments, but we do not support them currently.)
Applying a tactic to a goal can lead to a failure, when not all conditions are met, or is successful and produces a list of subgoals.
The goal is only proven successfully, if all its subgoals are proven.
In particular, if the goal is proven if the tactic application produces an empty list of subgoals.
We refer to tactic applications sometimes also as \emph{proof steps}.

We can think of proofs as trees, where goals are nodes and tactic applications are (hyper-)edges to other goals.
In a successful proof, all leaves are goals with a tactic application that produced an empty list of subgoals.

\subsection{Instrumentation to HOL Light}
In order to create a stable, well-defined environment, we fix a particular version of HOL Light with a pre-selected subset of tactics and a fixed library of basic theorems, which are proved in one well-defined order.
This is the ITP part of the environment which is written in OCaml with a few additional C++ functions.
Since it is non-trivial to find and build the exact correct set of libraries for this environment, we provide a prepackaged docker image.
It can be used as a reliable black box for proof search and as reinforcement learning environment, communicated with using a simple API.
We have also open sourced all the changes to the HOL Light system so that new modifications and forks are possible by third parties.

The prepackaged version we provide has the following additional instrumentation, which we describe below in detail:
\begin{itemize}
\item Logging of human-written proofs shipped with HOL Light.
\item A new API to interact with HOL Light for proof search.
\item Fast startup for distributed proof search.
\item A proof checker to remove the need to trust search algorithms.
\end{itemize}

\subsection{Proof Logging for Human Proofs}
We want to utilize the existing human proofs for both training and evaluation.
To that effect, we have instrumented the {\tt prove} method in HOL Light with extra logging code.
If HOL Light is executed in proof-dump mode, each invocation of the {\tt prove} function dumps the proven theorems and their proofs into files.
These \emph{proof logs} can then be converted to training examples (see Section~\ref{sec:training_examples}).

\subsection{Proof Assistant API}
The API provides two functions: (1) to apply tactics to goals and (2) to register theorems for future use in tactic applications.
Tactic applications are completely stateless and contain the goal, the tactic to be applied, and the tactic arguments.
The poof assistant (i.e. HOL Light in our implementation) returns the outcome of the tactic application, including the list of subgoals for successful applications.
The stateless tactic application interface frees us from the strict order on subgoals that HOL Light enforces in the human interface, and allows us to easily implement more advanced proof search strategies.

The tactic arguments can consist of a list of theorems.
Implemented naively, this list could make the tactic application request very large and could slow down the prover.
In the argument list of tactics we therefore allow theorems to be referenced by a fingerprint number.
The second API function allows us to register theorems such that HOL Light can resolve the fingerprints to theorems.
The registration of theorems is hence stateful, in contrast to tactic applications.

\subsection{Fast Startup}
Starting HOL Light and loading all the potentially needed libraries can take a long time - we measured it at up to 20 minutes.
This would be inhibitively long for proof search, especially in a distributed setting with thousands of workers and the startup time has to be paid for every worker.
The Proof Assistant API allows us to load only a minimal core of HOL Light and register the remaining theorems from the libraries using the API.
This brings the startup time of our HOL Light to mere seconds.

\subsection{Proof Checking}
Any bug in the implementation of a theorem prover could make its reasoning unsound, rendering the whole formalization effort futile.
For that reason, HOL Light is designed around a small trusted core of about 400 lines of OCaml code that builds proofs from few very basic rules.
OCaml's type system guarantees that a theorem object can only be constructed by this trusted core, and the rest of the HOL Light system can be seen as mere convenience features.

Our API allows researchers to implement proof search algorithms outside of OCaml.
The correctness of any proof found through the API thus relies on the correctness of our API implementation and the proof search itself.
We thus implemented a proof checker that avoids the need for trusting the proof search and even the API.
The proof checker compiles proofs into OCaml code that can be loaded in HOL Light, where they have to pass through the trusted core.
% Since the proofs that we want to check with the proof checker may depend on other theorems that could not be proven in the proof search, the proof checker hooks into the regular start-up of HOL Light and interleaves human-written proofs with the proofs to be checked.

\section{Benchmark}
\label{sec:benchmark}

We present three different corpora: ``core'', ``complex'', and ``flyspeck.''
The core corpus contains the basic theorems that needed to define the tactics and the complex corpus consists of theorems of complex calculus.
While proofs of core theorems are useful for training, we omit them in validation, since some tactics assume those theorems.
Flyspeck contains most of the lemmas and theorems of the Kepler conjecture.
% We used flyspeck only for testing generalization of a model trained on core and complex.
Together these three corpora encompass almost 30k theorems and proofs (see Table~\ref{tbl:corpora}).

We propose two tasks that can be measured on these benchmarks:
\begin{itemize}
\item Predict the tactic and tactic arguments that were employed in the human proof.
% \item Prove each of the subgoals in the proof log utilizing only some of the top level theorems preceding top level statement to which the subgoal belongs to.
\item Prove each of the theorems in the corpora while utilizing only those theorems as tactic arguments that also humans had available.
For that purpose, we provide all theorems in the three corpora in one unified list, in the order they were proven by humans.
\end{itemize}

\begin{table}[t]
\centering
\begin{tabular}{|llll|}
\hline
         & Definitions & Theorems & Proof states \\
core     & 239         &  2320    & 23512        \\
complex  & 398         &  16623   & 509621       \\
flyspeck & 1563        &  10519   & 538540       \\
\hline
all      & 2200        &  29462   & 1071673      \\
\hline
\end{tabular}\vspace{2mm}
\caption{The three corpora of the benchmark.}
\label{tbl:corpora}
\end{table}

\subsection{Training Examples}
\label{sec:training_examples}
Our training examples consist of a $\mathit{goal}$, a $\mathit{tactic}$, an $\mathit{arglist}$, and a $\mathit{negarglist}$.
The $\mathit{goal}$ is a provable statement, i.e. it is either a theorem from one of the corpora or a subgoal of a successful proof.
The $\mathit{tactic}$ is the ID of one of a preselected small set of tactics (currently consisting of 41 tactics) that led to a successful proof.
The $\mathit{arglist}$ is the list of theorems that were passed to a tactic application as arguments.
Additionally, there is a special argument signifying that the argument list was empty.

The $\mathit{negarglist}$ is an optional list of non-arguments that is not actually necessary for any proof.
$\mathit{negarglist}$ consists of high-scoring theorems that were not actually needed as arguments.
They are collected during proof search in our reinforcement learning pipeline, and the list is empty for all the examples generated from the human proof logs.

\subsection{Splits}
Before training and evaluation, we have split the top level theorems into three subsets:
training, validation and test set in a 60:20:20 ratio.
Since the goals occurring in the proof of a theorem are likely correlated with the theorem itself, we assign them the same split as the theorem.
The validation set can be used for continuous monitoring for proxy metrics of
the model during training.
The validation set is also occasionally used to measure
the end-to-end prover performance of the models during training.
The test set, on the other hand, must only be used extremely rarely for final
assessment of a few models before publishing a paper alongside their validation
set performance.

\subsection{Representation of Expressions}
All expressions are presented as S-expressions that have only few types of non-leaf nodes:
function applications, abstractions (i.e. lambda functions), variables, constants, and function types.
All other information, such as variable names, constant names, and type names, is given as leaf-nodes.
For example, the expression $f(x)$ for a function $f:\mathbb{R}\to\mathbb{R}$ looks as follows:
\texttt{(a (v (fun (real) real) f) (v real x))}.
These S-expressions have a unique correspondence to terms in HOL Light and are easy to parse into a tree.
However, our current models only observe the string version of these expressions.
Expressions are quite long in this representation: The average number of tokens in the goals is around 500, and the median is around 300.

For many operations, HOL Light automatically invents new generic types (e.g.~?345882) and generic variables (e.g. GEN\%PVAR\%9675) on the fly.
This leads to thousands of types and variables that often occur in only one (or a few) expressions, and hence would hardly get meaningful embeddings in typical deep learning approaches.
Further, tokens that are shared only between few expressions bear the risk of unintentionally giving away information about the relations between these statements.
We therefore decided to \emph{normalize} the data sets by mapping generic types and generic variables to a much smaller set of names while maintaining the semantics of all expressions.
After normalization, the number of distinct tokens is 1254.

% On the tactic side, in the human proof logs for flyspeck, the most common tactic
% is MESON\_TAC, in around 35\% of the proof steps. The number of tactic arguments
% is on average 1.5.
%TODO: Number of Tokens
%
%TODO: Median number of tokens, distribution
%
%TODO: Number of tactic arguments distribution
%
%TODO: Tactic frequency distribution

\section{DeepHOL Prover architecture}
\label{sec:deephol}

In this section, we describe the high-level architecture of our reference neural prover.
The intelligence is fully learned without any hand-crafted features, and with very simple data preprocessing.
In particular, we have not implemented any tweaks for the particular logic or interactive theorem prover (ITP).
All the engineering went into the neural network architecture, which is very generic, and into maintaining the proof search graph without any special regard for the particular ITP system.
In other words, DeepHOL currently uses HOL Light and its logic (HOL), but is not specialized to it.
We believe that our solution would also work with other goal-tactic based prover like Coq~\cite{bertot2013interactive}, HOL4~\cite{slind2008brief}, or Lean~\cite{de2015lean}.
Here we describe the details of our reference prover solution in detail.
% Fig.~\ref{fig:neuralprover} provides a high-level picture.

\subsection{Action Generator}
The most crucial part of our prover is the action generator that produces a list of tactic applications for a given goal.
We have split this into two subtasks:
\begin{itemize}
\item To rank the tactics, and
\item to create an argument list for each of the tactics (comprised of a list of theorems).
\end{itemize}
As noted earlier, DeepHOL is currently not using tactics that take arbitrary terms (formulas) as parameters.

For both subtasks, the action generator employs a neural network, which we describe in Section~\ref{ssec:neural}.
The ranking of tactic applications it produces is used in the proof search (Section~\ref{ssec:proofsearch}) to expand the proof search graph (Section~\ref{ssec:proofsearchgraph}).
% In a future section, we describe the precise details of the neural architecture that we used for predicting the tactics and their arguments.

% However our implementation merges identical subgoals, so we can end up with reconvergent paths: an acyclic graph in general.
% Such a graph is considered to be a successful proof graph if all its sinks are successfully proved by some tactic application.

%In the following subsection, we elaborate on the graph built during the proof search.

% directed acyclic graph of subgoals
%Each tactic application is represented by the set of outgoing arcs from each node.
%That is the result of each successful tactic application is a new set of subgoals to be proved.
%Note that it is possible that all of the attempted tactic applications fail: (for example, due to timeout) or does never change the subgoal up to variable renaming (i.e. alpha equivalence).
%In the trivial implementation this would result in a tree, however our implementation merges identical subgoals, so we can end up with reconvergent paths: an acyclic graph in general.
%Note however, that our implementation does not merge alpha equivalent subgoals, which might be useful as well.
%Such a graph is considered to be a successful proof graph if all its sinks are successfully proved by some tactic application.

\subsection{Proof Search Graph}
\label{ssec:proofsearchgraph}
The proof search graph is our data structure that captures the state of the proof search, and allows us to detect when a proof for the original goal is available.
The nodes of the proof search graph are the goals that we have seen in the proof search, including the original goal statement that we want to prove.
Each goal can have multiple alternative tactic applications, each of which might result in multiple subgoals.
That is, tactic applications are labelled hyperedges in the proof search graph.

The proof search graph provides some features that allow us to prune some subtrees of the search early:
First, whenever a tactic application closes a subgoal, this information is traced back to the parent
subgoals and each alternate tactic application (and its whole sub-branch) is marked as closed is discarded from the queue to be processed.
If during this recursive process the root node is reached, then the proof is closed and the proof process stops.
Second, when all tactic applications for a goal fail we mark that goal as unsuccessful.
Similar to tracing closed goals, the proof search graph automatically traces the siblings of unsuccessful subgoals that become superflous, and mark them unsuccessful as well.
Third, when tactic applications produce identical subgoals, we let them point to the same node in the proof search graph. We refer to this as \emph{subgoal sharing}, and once a subgoal is newly shared, previously stored information about subgoals being closed or ignored is be propagated through the search graph.

\subsection{Proof Search}
\label{ssec:proofsearch}
Our proof search is a simple breadth first search.
In each iteration, its expands all leaf nodes (i.e. goals that have not been expanded yet).
To exand a goal, it calls the action generator to generate a list of tactic applications, and applies them in order.
It stops applying tactics to a goal, when it reaches a maximum number of unsuccessful tactic applications or a minimum number of successful tactic applications.
Whenever a complete proof is found for the top level goal, the proof search is stopped and the whole proof search graph is serialized and stored as the result.
Also, the proof search finishes if the search graph reaches a prescribed limit on the number of subgoals or the proof search times out.

Note that subgoal sharing, as explained in Section~\ref{ssec:proofsearchgraph}, is crucial for our proof search:
Without subgoal sharing the search process could end up oscillating between two formulas by rewriting the same subterm back and forth using the same equation.

\subsection{Neural Architectures}
\label{ssec:neural}
\begin{figure}
\begin{center}
\includegraphics[width=\columnwidth,clip,trim={3cm 5cm 3cm 5cm}]{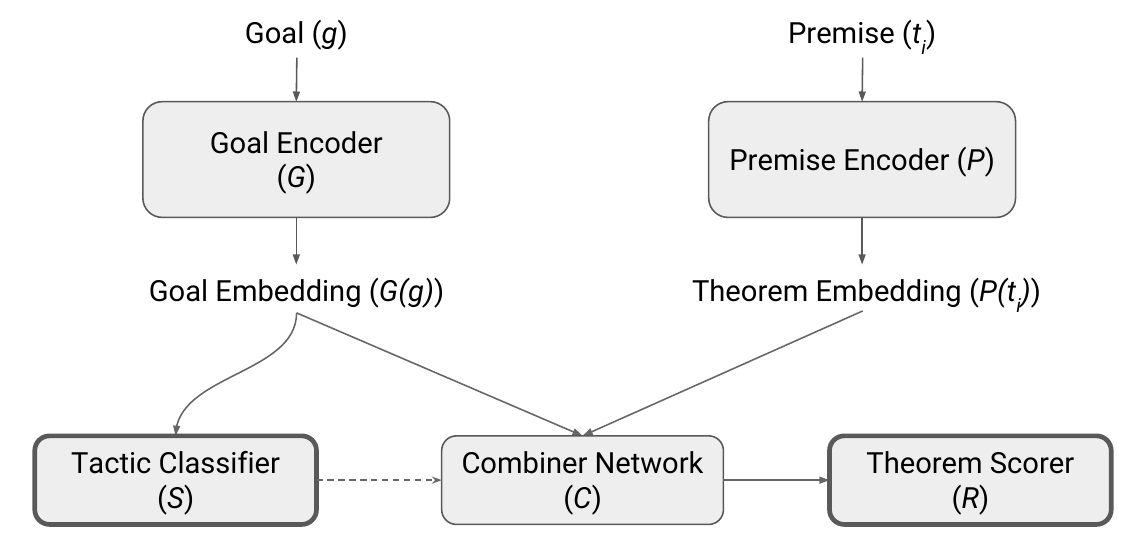}
\end{center}
\caption{Two-tower neural architecture for ranking actions.}
\label{fig:modelarchitecture}
\end{figure}
For the generation and ranking of actions in the action generator, we use a deep, two-tower neural network depicted in Figure~\ref{fig:modelarchitecture}.
The predictions of the neural network are based on a single goal, represented as an S-expression of the HOL Light term (i.e. a string).
(In HOL Light, each goal consists of a list of hypotheses and a conclusion, and we currently drop the hypotheses before we feed a goal to the neural network.)

The neural network has two separate prediction heads $S$ and $R$.
The goal tower $G$ computes an embedding $G(g)$ of the current goal $g$ and infers a scoring vector $S(G(g))$ for the fixed set of tactics where the tactic classifier $S$ is a linear layer producing logits of a softmax classifier.
The premise tower $P$ computes a fixed size embedding $P(t_i)$ of all possible tactic arguments $t_i$ in the scope of the goal to be proved.
The ranking of the premises is performed by a combiner network $C$ that takes the concatenation of the goal embedding, the premise embedding and possibly that of the tactic $T_j$ to be applied: $r(t_i)=C(G(g), P(t_i), T_j)$,
where $r(t_i)$ is the score of theorem $t_i$ for its being a useful tactic argument
for transforming the current goal $g$ towards a closed proof.
We have also tried the unconditioned setup, in which the ranking of the tactic
arguments is independent of that of the tactic to be applied, that is $r(t_i)=C(G(g), P(t_i))$.
In essence, we propose a hybrid architecture that both predicts the correct
tactic to be applied, as well as rank the premise parameters required for meaningful application
of the tactics.

\subsection{Supervised Learning}
We started training DeepHOL in a supervised learning setup, for which we use the human proof logs.
We have split our data into test, train, and validation set on the theorem level, as described in Section~\ref{sec:benchmark}.
We always report both validation and test set performance for the final result
to verify that we did not over-fit on the validation set.
Continuous measurements and ablation analyses are reported only on the validation
or training set.

\subsection{Reinforcement Learning Loop}
In the reinforcement learning loop, we have both a trainer and multiple
provers running continuously.
The training is (optionally) seeded with training examples from existing
(human/generated) proof logs.
Then, we run the neural prover in \emph{rounds}, each round trying to prove a
random sample of theorems in the training set.
Training examples extracted from successful proof logs of each round of our neural prover
are mixed in continuously.
Training examples of more recent rounds (\emph{fresh} examples) can be weighed
differently from older rounds (\emph{historical} examples) during the training process.

To summarize, our loop works with the following four kinds of training example pools:
\begin{enumerate}
\item (optional) Human training examples as seed.
\item (optional) Inherited computer generated examples as seed: in addition to using human
  training examples as seed, examples generated during any previous experiments with our prover
  can also be used as seed.
  In our current experiments, we used examples that were generated by a prover that was run on
  the whole training set utilizing a model that was trained in purely supervised manner.
\item Fresh generated loop examples (examples that were produced in the last $k$ rounds, where $k$ is a user-settable parameter).
\item Historical training loop examples (examples that were produced in all but the last $k$ rounds).
\end{enumerate}
During training, batches are filled with examples from each pool according to a prescribed split ratio.
This means that the ratio of different kinds of examples the model is trained on does not shift as more examples are generated by the loop.
Most importantly, it also ensures that examples from freshly constructed new proofs
show up quickly and deterministically during the training process.
Note that although we can make use of human and inherited proof traces, the system can learn without any supervision or initial seed data.
However, preliminary experiments have shown that, in its current form, it learns
inferior models compared to those that were seeded with human proofs.

\subsubsection{Proof Pruning}
The argument lists of tactic applications in the reinforcement learning loop are quite long,
and they contain superfluous elements.
In order to obtain high quality training data for tactic
argument prediction, we \emph{prune} the parameter list before using them for training.
For all tactics that take a list of theorems as an argument, our current
implementation generates a list of fixed length.
For successful tactic applications, we then iterate over the arguments in
reverse score order and greedily omit those arguments that do not change the
outcome of the tactic application.
While a non-greedy approach might yield even shorter argument lists, it would also
take longer to compute.
In practice, our approach produces short argument lists with minimal effort.
Removed parameters are stored as ``hard negatives'' and utilized during training.
\begin{table}
\centering
  \begin{tabular}{|l|c|}
    \hline
    Description & Proof success \\
    \hline \hline
ASM\_MESON\_TAC                         &  $6.1$\%   \\ \hline
ASM\_MESON\_TAC + & \\
argument selection    &  $9.2$\%   \\ \hline\hline
% 1 Convolutional layer                  &  $24.1$\%      \\ \hline
% 2 Convolutional layers                 &  $17.1$\%      \\ \hline
WaveNet                  &  $31.72$\%      \\ \hline
Deeper WaveNet           &  $32.65$\%        \\ \hline
Wider WaveNet            &  $27.60$\%      \\ \hline
% Wider WaveNet                 &  $17.7$\%      \\ \hline
% Wider WaveNet, regularized     &  $27.2$\%     \\ \hline
% Thin WaveNet, tactic dependent &  $22.5$\%     \\ \hline
% Thin WaveNet, tactic dependent & \\
% + historical &  $32.8$\%                      \\ \hline
\hline
Loop                           & $36.3$\%     \\ \hline
Trained on loop output         & $36.8$\%     \\ \hline
Loop tactic dependent          & $\mathbf{38.9}$\%    \\ \hline
% Loop on subgoals               & $34.6$\%     \\ \hline
\end{tabular}
\vspace{2mm}
  \caption{Percentage of theorems closed using various models
    on the validation set of the \emph{complex} corpus comprising of 3225 theorems.
    First two lines are trivial baselines that call HOL Light's built-in first
    order theorem prover with and without utilizing our argument selection model.
    The middle section shows results of models trained in a supervised scenario on
    human proofs.
    The last four lines report results using our reinforcement learning loops.
    % Test set performance is evaluated on two models that yielded best result
    % on the validation set in their category.
    \label{results}
  }
\end{table}

\section{Results and Comparisons}
\label{sec:comparisons}
In this section, we first present several baseline results based on imitation (i.e. fully supervised) learning.
Then we come to our reinforcement learning results using a WaveNet~\cite{van2016wavenet} based encoder
architecture, but with three different training methodologies.

\subsection{Model Training Hyperparameters}
All models were trained with the Adam optimizer~\cite{kingma2014adam} and
exponentially decreasing learning rate starting at $0.0001$ with decay
rate $0.98$ at every $100000$ steps.
For evaluation, we use moving exponential parameter averaging at a rate of
$0.9999$ per step~\cite{polyak1990new,polyak1992acceleration}.
First, we established trivial baselines by running the built-in first-order
theorem prover {\tt ASM\_MESON\_TAC} on each theorem on the dataset with empty argument list and
with an argument list predicted with our baseline WaveNet model.
Next, we compare the performance of various WaveNet style architectures.
Finally, we report our reinforcement learning experiments on the
complex analysis corpus.
Our final prover performance numbers are summarized in Table~\ref{results}.

\subsection{Comparison of Model Architectures}
We trained and evaluated a large number of networks and present a sample of our findings.
% with
% up to four layers, rectified linear application, stride $2$ from the second
% layer, patch size in $\{3, 5, 7\}$,
% and filter size in $\{128, 256, 512, 1024, 2048\}$.
During our experiments, we looked at the following proxy metrics:
\begin{enumerate}
\item Accuracy of tactic prediction out of the $41$ possible tactics.
  (Ranging between $38$\% and $42$\% for most models.)
\item Success rate of selecting a positive tactic argument over a randomly selected negative argument. (Around 1\% error rate).
\end{enumerate}
For the encoders, we have tried WaveNet~\cite{van2016wavenet} style networks
with different hyper-parameters.
The various results on the complex analysis corpus are based on imitation learning
and the combination of imitation learning and reinforcement learning.
In the base model we used two WaveNet blocks of four layers each. The number of filters in
each block was either $128$ or $256$. As one can see in Table~\ref{results},
the network with less filters did better. Then we tried a deeper variant
with four blocks of five layers each, in this case with depth $128$.
Here the deeper network with more blocks, which has $47$ million parameters, turned out to be superior.
Both architectures incorporate fully connected combiner layers with
additional dropout layers before each of them. The ratio of dropped out neurons
during training was $0.3$.
Note that the reinforcement learning experiments was performed earlier and
was ran with the narrow architecture (with $128$ filters in each layer) and
with two wavenet blocks.

\subsection{Reinforcement Learning}
\begin{figure}
  \includegraphics[width=\linewidth]{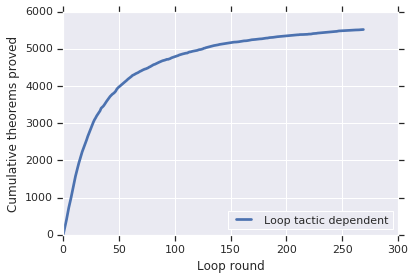}
  \vspace{-6mm}
  \caption{This figure presents the cumulative number of proofs closed by the tactic dependent loop.
  The total number of theorems in the training set is 10199.}
  \label{fig:cumulative_tac_depend}
\end{figure}
In our reinforcement learning set up, the model training runs on a single GPU,
while theorem proving is performed in a distributed manner: we attempt to prove 2000
randomly selected theorems from the training set of the union of the
complex and core corpora in every round.
At the start of each round, we fetch the latest trained model checkpoint and precompute
the theorem argument embedding for each theorem in the complex and core libraries.
This precomputation greatly accelerates the ranking of the tactic arguments.
The proof search is distributed over 1000 cores and we set a computation limit of
100 explored proof states and a total timeout of 300 seconds.
Each individual tactic application has a timeout of 5 seconds.
% arxiv
% (most tactics succeed or fail within a fraction of a second, this timeout is only relevant for long running tactics).
Additionally, for each example, we pick prover options uniformly
in the ranges described by Table~\ref{table:params}, to increase the diversity of
the generated proofs.
This also increases the chance of finding a proof at all for harder statements.
\begin{table}[t]
\centering
  \begin{tabular}{|l|c|}
    \hline
    Maximum number of top tactics explored & $[6, 16]$ \\ \hline
    Maximum {\it successful} tactic applications & $[3, 6]$ \\ \hline
    Number of selected tactic arguments & [$1$, $32$] \\ \hline
  \end{tabular}\vspace{2mm}
  \caption{Randomized proof search parameters and their ranges.}
  \label{table:params}
\end{table}
\begin{table}[t]
\centering
  \begin{tabular}{|l|c|}
    \hline
    & Theorems proved \\
    Name &(\% of training set) \\ \hline
    Loop & $5679$ ($55.7$\%) \\
    Loop tactic dependent & $5518$ ($54.1$\%) \\
    Loop on subgoals & $1988$ ($19.5$\%) \\ \hline
    \textbf{Union} & $5919$ ($58.0$\%) \\ \hline
  \end{tabular}\vspace{2mm}
  \caption{Total count of proofs found by each loop.}
  \label{table:cumulative_proofs}
\end{table}
\begin{figure}
  \includegraphics[width=\linewidth]{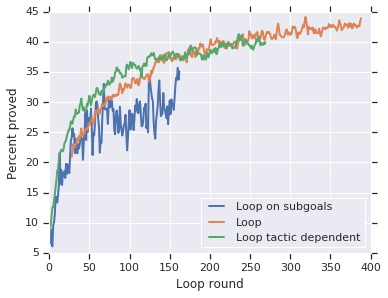}
  \vspace{-6mm}
  \caption{Percentage of theorems proved in each round of the loop.
  Each round samples 2000 theorems from the training set.}
  \label{fig:percent_proved}
  \vspace{-2mm}
\end{figure}

Given the high computational cost of running the reinforcement learning loop,
we have only tried a couple of variants. Each of our these experiments
use the same version of WaveNet~\cite{van2016wavenet}
architecture (with $128$ filters in each layer).
In our first loop experiment, ``Loop'', we trained a loop with tactic
independent argument selection. That is, the tactic argument ranking was
independent of the tactic chosen, and we pick only top level theorems
to be proved by proof search in the reinforcement learning scenario.
Alongside our first loop, we trained a separate model ``Trained on loop output'' that was not used in the loop for proof search guidance, but did benefit from a curriculum-style learning, since it trained in parallel to the loop.
%arxiv
%This model, ``Trained on loop output'', does very well, even better that other models that are trained purely on the aggregate of the loop output.
%In future work, we hope to run additional experiments to explore the contributions of curricula learning on synthetic data vs. training the model in a reinforcement loop.
In our second loop experiment, ``Loop tactic dependent'', we have trained a model in which the arguments ranking
depends on the selected tactic.
In our third loop experiment, ``Loop on subgoals'', the proof search can pick from any of the internal
proof states from the training set of the joined core+complex corpus.
This was motivated partially by the success of \cite{zombori2019curriculum},
we tried to run a reinforcement learning loop in which we train for
solving each subgoal separately, hoping that it will help for learning longer proofs.
This means, that we expected a bigger variety of theorems to be generated during
proof search. However, our naive implementation did not seem to end up with improved results.
Performance of each loop's final checkpoint on the validation set is presented in Table~\ref{results}.
We also ran the final checkpoint of the ``Loop'' on a sample of 2000 proofs from the {\it flyspeck} dataset; we closed 752 (37.0\%) of these proofs automatically.

While it was too computationally expensive to track the validation performance on every round of the loop, we did record the performance on the training set.
In Fig.~\ref{fig:cumulative_tac_depend}, we show the cumulative number of proofs closed by the tactic dependent loop at each round.
% We expect many of the ``easy'' proofs to be closed in the earliest rounds, and in later After several rounds, we find fewer new proofs, however we continue to find
Recall that in each round we sample theorems from the training set and use the most recent checkpoint to guide the proof search.
In Fig.~\ref{fig:percent_proved}, we show the percentage of the sampled theorems that are proved in each round.

\section{Conclusion}
\label{sec:conclusion}

We presented a machine learning oriented open source environment for higher-order theorem proving as well as a neural network based automated prover, trained on a large-scale reinforcement learning system.
We also suggest a benchmark for machine reasoning in higher-order logic on a relatively large and practically relevant corpus of theorems with varying complexity.
Our benchmark includes purely neural network based baselines, which demonstrate strong automated reasoning capabilities, including premise selection from a large number of theorems.
We hope that our initial effort fosters collaboration and paves the way for strong and practical AI systems that can learn to reason efficiently in large formal theories.

\section*{Acknowledgements}
We would like to thank Alex Alemi, Geoffrey Irving, Cezary Kaliszyk, Thibault Gauthier, Ramana Kumar, Viktor Toman, and Josef Urban for their insightful comments and contributions to early versions of this work.

\bibliography{holist}

\begin{thebibliography}{50}
\providecommand{\natexlab}[1]{#1}
\providecommand{\url}[1]{\texttt{#1}}
\expandafter\ifx\csname urlstyle\endcsname\relax
  \providecommand{\doi}[1]{doi: #1}\else
  \providecommand{\doi}{doi: \begingroup \urlstyle{rm}\Url}\fi

\bibitem[Panayotov et~al.(2015)Panayotov, Chen, Povey, and
  Khudanpur]{panayotov2015librispeech}
Vassil Panayotov, Guoguo Chen, Daniel Povey, and Sanjeev Khudanpur.
\newblock Librispeech: an asr corpus based on public domain audio books.
\newblock In \emph{Acoustics, Speech and Signal Processing (ICASSP), 2015 IEEE
  International Conference on}, pages 5206--5210. IEEE, 2015.

\bibitem[Bennett et~al.(2007)Bennett, Lanning, et~al.]{bennett2007netflix}
James Bennett, Stan Lanning, et~al.
\newblock The netflix prize.
\newblock In \emph{Proceedings of KDD cup and workshop}, volume 2007, page~35.
  New York, NY, USA, 2007.

\bibitem[Deng et~al.(2009)Deng, Dong, Socher, Li, Li, and
  Fei-Fei]{deng2009imagenet}
Jia Deng, Wei Dong, Richard Socher, Li-Jia Li, Kai Li, and Li~Fei-Fei.
\newblock Imagenet: A large-scale hierarchical image database.
\newblock In \emph{Computer Vision and Pattern Recognition, 2009. CVPR 2009.
  IEEE Conference on}, pages 248--255. Ieee, 2009.

\bibitem[Lin et~al.(2014)Lin, Maire, Belongie, Hays, Perona, Ramanan,
  Doll{\'a}r, and Zitnick]{lin2014microsoft}
Tsung-Yi Lin, Michael Maire, Serge Belongie, James Hays, Pietro Perona, Deva
  Ramanan, Piotr Doll{\'a}r, and C~Lawrence Zitnick.
\newblock Microsoft coco: Common objects in context.
\newblock In \emph{European conference on computer vision}, pages 740--755.
  Springer, 2014.

\bibitem[Bojar et~al.(2014)Bojar, Buck, Federmann, Haddow, Koehn, Leveling,
  Monz, Pecina, Post, Saint-Amand, et~al.]{bojar2014findings}
Ondrej Bojar, Christian Buck, Christian Federmann, Barry Haddow, Philipp Koehn,
  Johannes Leveling, Christof Monz, Pavel Pecina, Matt Post, Herve Saint-Amand,
  et~al.
\newblock Findings of the 2014 workshop on statistical machine translation.
\newblock In \emph{Proceedings of the ninth workshop on statistical machine
  translation}, pages 12--58, 2014.

\bibitem[Rajpurkar et~al.(2016)Rajpurkar, Zhang, Lopyrev, and
  Liang]{rajpurkar2016squad}
Pranav Rajpurkar, Jian Zhang, Konstantin Lopyrev, and Percy Liang.
\newblock Squad: 100,000+ questions for machine comprehension of text.
\newblock \emph{arXiv preprint arXiv:1606.05250}, 2016.

\bibitem[Hales et~al.(2017)Hales, Adams, Bauer, Dang, Harrison, Le~Truong,
  Kaliszyk, Magron, McLaughlin, Nguyen, et~al.]{hales2017formal}
Thomas Hales, Mark Adams, Gertrud Bauer, Tat~Dat Dang, John Harrison, Hoang
  Le~Truong, Cezary Kaliszyk, Victor Magron, Sean McLaughlin, Tat~Thang Nguyen,
  et~al.
\newblock A formal proof of the kepler conjecture.
\newblock In \emph{Forum of Mathematics, Pi}, volume~5. Cambridge University
  Press, 2017.

\bibitem[Urban et~al.(2008)Urban, Sutcliffe, Pudl{\'a}k, and
  Vysko{\v{c}}il]{urban2008malarea}
Josef Urban, Geoff Sutcliffe, Petr Pudl{\'a}k, and Ji{\v{r}}{\'\i}
  Vysko{\v{c}}il.
\newblock Malarea sg1-machine learner for automated reasoning with semantic
  guidance.
\newblock In \emph{International Joint Conference on Automated Reasoning},
  pages 441--456. Springer, 2008.

\bibitem[Gauthier et~al.(2017)Gauthier, Kaliszyk, and
  Urban]{gauthier2017tactictoe}
Thibault Gauthier, Cezary Kaliszyk, and Josef Urban.
\newblock Tactictoe: Learning to reason with hol4 tactics.
\newblock In \emph{LPAR-21. 21st International Conference on Logic for
  Programming, Artificial Intelligence and Reasoning}, volume~46, pages
  125--143, 2017.

\bibitem[Huang et~al.(2018)Huang, Dhariwal, Song, and
  Sutskever]{huang2018gamepad}
Daniel Huang, Prafulla Dhariwal, Dawn Song, and Ilya Sutskever.
\newblock Gamepad: A learning environment for theorem proving.
\newblock \emph{arXiv preprint arXiv:1806.00608}, 2018.

\bibitem[Slind and Norrish(2008)]{slind2008brief}
Konrad Slind and Michael Norrish.
\newblock A brief overview of hol4.
\newblock In \emph{International Conference on Theorem Proving in Higher Order
  Logics}, pages 28--32. Springer, 2008.

\bibitem[Bertot and Cast{\'e}ran(2013)]{bertot2013interactive}
Yves Bertot and Pierre Cast{\'e}ran.
\newblock \emph{Interactive theorem proving and program development: Coq’Art:
  the calculus of inductive constructions}.
\newblock Springer Science \& Business Media, 2013.

\bibitem[Mizar()]{mml-homepage}
Mizar.
\newblock The {Mizar} {Mathematical} {Library}.
\newblock URL \url{http://mizar.org}.
\newblock Accessed: 2018/01/18.

\bibitem[Wenzel et~al.(2008)Wenzel, Paulson, and Nipkow]{wenzel08isabelle}
Makarius Wenzel, Lawrence~C. Paulson, and Tobias Nipkow.
\newblock The isabelle framework.
\newblock In Otmane~A{\"{\i}}t Mohamed, C{\'{e}}sar~A. Mu{\~{n}}oz, and
  Sofi{\`{e}}ne Tahar, editors, \emph{Theorem Proving in Higher Order Logics,
  21st International Conference, TPHOLs 2008, Montreal, Canada, August 18-21,
  2008. Proceedings}, volume 5170 of \emph{Lecture Notes in Computer Science},
  pages 33--38. Springer, 2008.

\bibitem[de~Moura et~al.(2015)de~Moura, Kong, Avigad, Van~Doorn, and von
  Raumer]{de2015lean}
Leonardo de~Moura, Soonho Kong, Jeremy Avigad, Floris Van~Doorn, and Jakob von
  Raumer.
\newblock The lean theorem prover (system description).
\newblock In \emph{International Conference on Automated Deduction}, pages
  378--388. Springer, 2015.

\bibitem[Klein et~al.(2009)Klein, Elphinstone, Heiser, Andronick, Cock, Derrin,
  Elkaduwe, Engelhardt, Kolanski, Norrish, Sewell, Tuch, and
  Winwood]{Klein2009seL4}
Gerwin Klein, Kevin Elphinstone, Gernot Heiser, June Andronick, David Cock,
  Philip Derrin, Dhammika Elkaduwe, Kai Engelhardt, Rafal Kolanski, Michael
  Norrish, Thomas Sewell, Harvey Tuch, and Simon Winwood.
\newblock sel4: Formal verification of an os kernel.
\newblock In \emph{Proceedings of the ACM SIGOPS 22Nd Symposium on Operating
  Systems Principles}, SOSP '09, pages 207--220, New York, NY, USA, 2009. ACM.
\newblock ISBN 978-1-60558-752-3.
\newblock \doi{10.1145/1629575.1629596}.
\newblock URL \url{http://doi.acm.org/10.1145/1629575.1629596}.

\bibitem[Kaliszyk and Urban(2014)]{kaliszyk2014learning}
Cezary Kaliszyk and Josef Urban.
\newblock Learning-assisted automated reasoning with flyspeck.
\newblock \emph{Journal of Automated Reasoning}, 53\penalty0 (2):\penalty0
  173--213, 2014.

\bibitem[Kaliszyk and Urban(2012)]{kaliszyk2012initial}
Cezary Kaliszyk and Josef Urban.
\newblock Initial experiments with external provers and premise selection on
  hol light corpora.
\newblock 2012.

\bibitem[Alemi et~al.(2016)Alemi, Chollet, Irving, E{\'e}n, Szegedy, and
  Urban]{alemi2016deepmath}
Alexander~A Alemi, Fran{\c{c}}ois Chollet, Geoffrey Irving, Niklas E{\'e}n,
  Christian Szegedy, and Josef Urban.
\newblock Deepmath-deep sequence models for premise selection.
\newblock In \emph{Advances in Neural Information Processing Systems}, pages
  2235--2243, 2016.

\bibitem[Schulz(2002)]{Sch02-AICOMM}
Stephan Schulz.
\newblock {E - A Brainiac Theorem Prover}.
\newblock \emph{AI Commun.}, 15\penalty0 (2-3):\penalty0 111--126, 2002.

\bibitem[Whalen(2016)]{whalen2016holophrasm}
Daniel Whalen.
\newblock Holophrasm: a neural automated theorem prover for higher-order logic.
\newblock \emph{arXiv preprint arXiv:1608.02644}, 2016.

\bibitem[Megill(1997)]{megill1997metamath}
Norman Megill.
\newblock Metamath: A computer language for pure mathematics.
\newblock 1997.

\bibitem[Loos et~al.(2017)Loos, Irving, Szegedy, and Kaliszyk]{loos2017deep}
Sarah Loos, Geoffrey Irving, Christian Szegedy, and Cezary Kaliszyk.
\newblock Deep network guided proof search.
\newblock \emph{arXiv preprint arXiv:1701.06972}, 2017.

\bibitem[Kaliszyk et~al.(2017)Kaliszyk, Chollet, and
  Szegedy]{kaliszyk2017holstep}
Cezary Kaliszyk, Fran{\c{c}}ois Chollet, and Christian Szegedy.
\newblock Holstep: A machine learning dataset for higher-order logic theorem
  proving.
\newblock \emph{arXiv preprint arXiv:1703.00426}, 2017.

\bibitem[Kaliszyk et~al.(2018)Kaliszyk, Urban, Michalewski, and
  Ol{\v{s}}{\'a}k]{kaliszyk2018reinforcement}
Cezary Kaliszyk, Josef Urban, Henryk Michalewski, and Mirek Ol{\v{s}}{\'a}k.
\newblock Reinforcement learning of theorem proving.
\newblock \emph{arXiv preprint arXiv:1805.07563}, 2018.

\bibitem[Fan et~al.(2008)Fan, Chang, Hsieh, Wang, and Lin]{fan2008liblinear}
Rong-En Fan, Kai-Wei Chang, Cho-Jui Hsieh, Xiang-Rui Wang, and Chih-Jen Lin.
\newblock Liblinear: A library for large linear classification.
\newblock \emph{Journal of machine learning research}, 9\penalty0
  (Aug):\penalty0 1871--1874, 2008.

\bibitem[Otten and Bibel(2003)]{otten2003leancop}
Jens Otten and Wolfgang Bibel.
\newblock {leanCoP:} lean connection-based theorem proving.
\newblock \emph{J. Symb. Comput.}, 36\penalty0 (1-2):\penalty0 139--161, 2003.

\bibitem[Schulz(2000)]{Sch00}
Stephan Schulz.
\newblock \emph{{Learning search control knowledge for equational deduction}},
  volume 230 of \emph{DISKI}.
\newblock Infix Akademische Verlagsgesellschaft, 2000.
\newblock ISBN 978-3-89838-230-4.

\bibitem[Duncan et~al.(2004)Duncan, Bundy, Levine, Storkey, and
  Pollet]{duncan2004use}
Hazel Duncan, A~Bundy, J~Levine, A~Storkey, and M~Pollet.
\newblock The use of data-mining for the automatic formation of tactics.
\newblock 2004.

\bibitem[Urban et~al.(2011)Urban, Vysko{\v{c}}il, and
  {\v{S}}t{\v{e}}p{\'a}nek]{urban2011malecop}
Josef Urban, Ji{\v{r}}{\'\i} Vysko{\v{c}}il, and Petr {\v{S}}t{\v{e}}p{\'a}nek.
\newblock Malecop machine learning connection prover.
\newblock In \emph{International Conference on Automated Reasoning with
  Analytic Tableaux and Related Methods}, pages 263--277. Springer, 2011.

\bibitem[K{\"u}hlwein et~al.(2012)K{\"u}hlwein, van Laarhoven, Tsivtsivadze,
  Urban, and Heskes]{kuhlwein2012overview}
Daniel K{\"u}hlwein, Twan van Laarhoven, Evgeni Tsivtsivadze, Josef Urban, and
  Tom Heskes.
\newblock Overview and evaluation of premise selection techniques for large
  theory mathematics.
\newblock In \emph{International Joint Conference on Automated Reasoning},
  pages 378--392. Springer, 2012.

\bibitem[Kaliszyk and Urban(2013)]{kaliszyk2013stronger}
Cezary Kaliszyk and Josef Urban.
\newblock Stronger automation for flyspeck by feature weighting and strategy
  evolution.
\newblock 2013.

\bibitem[K{\"u}hlwein et~al.(2013)K{\"u}hlwein, Blanchette, Kaliszyk, and
  Urban]{kuhlwein2013mash}
Daniel K{\"u}hlwein, Jasmin~Christian Blanchette, Cezary Kaliszyk, and Josef
  Urban.
\newblock Mash: machine learning for sledgehammer.
\newblock In \emph{International Conference on Interactive Theorem Proving},
  pages 35--50. Springer, 2013.

\bibitem[Alama et~al.(2014)Alama, Heskes, K{\"u}hlwein, Tsivtsivadze, and
  Urban]{alama2014premise}
Jesse Alama, Tom Heskes, Daniel K{\"u}hlwein, Evgeni Tsivtsivadze, and Josef
  Urban.
\newblock Premise selection for mathematics by corpus analysis and kernel
  methods.
\newblock \emph{Journal of Automated Reasoning}, 52\penalty0 (2):\penalty0
  191--213, 2014.

\bibitem[Bridge et~al.(2014)Bridge, Holden, and Paulson]{BridgeHP14}
James~P. Bridge, Sean~B. Holden, and Lawrence~C. Paulson.
\newblock Machine learning for first-order theorem proving.
\newblock \emph{J. Autom. Reasoning}, pages 1--32, 2014.
\newblock ISSN 0168-7433.
\newblock \doi{10.1007/s10817-014-9301-5}.
\newblock URL \url{http://dx.doi.org/10.1007/s10817-014-9301-5}.

\bibitem[Kaliszyk et~al.(2014{\natexlab{a}})Kaliszyk, Urban, and
  Vysko{\v{c}}il]{kaliszyk2014machinelearner}
Cezary Kaliszyk, Josef Urban, and Ji{\v{r}}{\'\i} Vysko{\v{c}}il.
\newblock Machine learner for automated reasoning 0.4 and 0.5.
\newblock \emph{arXiv preprint arXiv:1402.2359}, 2014{\natexlab{a}}.

\bibitem[Kaliszyk et~al.(2014{\natexlab{b}})Kaliszyk, Mamane, and
  Urban]{kaliszyk2014machine}
Cezary Kaliszyk, Lionel Mamane, and Josef Urban.
\newblock Machine learning of coq proof guidance: First experiments.
\newblock \emph{arXiv preprint arXiv:1410.5467}, 2014{\natexlab{b}}.

\bibitem[F{\"a}rber and Kaliszyk(2015)]{farber2015random}
Michael F{\"a}rber and Cezary Kaliszyk.
\newblock Random forests for premise selection.
\newblock In \emph{International Symposium on Frontiers of Combining Systems},
  pages 325--340. Springer, 2015.

\bibitem[Kaliszyk and Urban(2015{\natexlab{a}})]{kaliszyk2015mizar}
Cezary Kaliszyk and Josef Urban.
\newblock Mizar 40 for mizar 40.
\newblock \emph{Journal of Automated Reasoning}, 55\penalty0 (3):\penalty0
  245--256, 2015{\natexlab{a}}.

\bibitem[Kaliszyk et~al.(2015)Kaliszyk, Urban, and
  Vyskocil]{kaliszyk2015efficient}
Cezary Kaliszyk, Josef Urban, and Jir{\'i} Vyskocil.
\newblock Efficient semantic features for automated reasoning over large
  theories.
\newblock In \emph{IJCAI}, 2015.

\bibitem[Kaliszyk and Urban(2015{\natexlab{b}})]{kaliszyk2015femalecop}
Cezary Kaliszyk and Josef Urban.
\newblock Femalecop: Fairly efficient machine learning connection prover.
\newblock In \emph{Logic for Programming, Artificial Intelligence, and
  Reasoning}, pages 88--96. Springer, 2015{\natexlab{b}}.

\bibitem[Kaliszyk and Urban(2015{\natexlab{c}})]{kaliszyk2015learning}
Cezary Kaliszyk and Josef Urban.
\newblock Learning-assisted theorem proving with millions of lemmas.
\newblock \emph{Journal of symbolic computation}, 69:\penalty0 109--128,
  2015{\natexlab{c}}.

\bibitem[Gauthier and Kaliszyk(2015)]{gauthier2015premise}
Thibault Gauthier and Cezary Kaliszyk.
\newblock Premise selection and external provers for hol4.
\newblock In \emph{Proceedings of the 2015 Conference on Certified Programs and
  Proofs}, pages 49--57. ACM, 2015.

\bibitem[Blanchette et~al.(2016)Blanchette, Greenaway, Kaliszyk, K{\"u}hlwein,
  and Urban]{blanchette2016learning}
Jasmin~Christian Blanchette, David Greenaway, Cezary Kaliszyk, Daniel
  K{\"u}hlwein, and Josef Urban.
\newblock A learning-based fact selector for isabelle/hol.
\newblock \emph{Journal of Automated Reasoning}, 57\penalty0 (3):\penalty0
  219--244, 2016.

\bibitem[Wang et~al.(2017)Wang, Tang, Wang, and Deng]{wang2017premise}
Mingzhe Wang, Yihe Tang, Jian Wang, and Jia Deng.
\newblock Premise selection for theorem proving by deep graph embedding.
\newblock In \emph{Advances in Neural Information Processing Systems}, pages
  2786--2796, 2017.

\bibitem[Van Den~Oord et~al.(2016)Van Den~Oord, Dieleman, Zen, Simonyan,
  Vinyals, Graves, Kalchbrenner, Senior, and Kavukcuoglu]{van2016wavenet}
A{\"a}ron Van Den~Oord, Sander Dieleman, Heiga Zen, Karen Simonyan, Oriol
  Vinyals, Alex Graves, Nal Kalchbrenner, Andrew Senior, and Koray Kavukcuoglu.
\newblock Wavenet: A generative model for raw audio.
\newblock \emph{CoRR abs/1609.03499}, 2016.

\bibitem[Kingma and Ba(2014)]{kingma2014adam}
Diederik~P Kingma and Jimmy Ba.
\newblock Adam: A method for stochastic optimization.
\newblock \emph{arXiv preprint arXiv:1412.6980}, 2014.

\bibitem[Polyak(1990)]{polyak1990new}
Boris~Teodorovich Polyak.
\newblock A new method of stochastic approximation type.
\newblock \emph{Avtomatika i telemekhanika}, 7:\penalty0 98--107, 1990.

\bibitem[Polyak and Juditsky(1992)]{polyak1992acceleration}
Boris~T Polyak and Anatoli~B Juditsky.
\newblock Acceleration of stochastic approximation by averaging.
\newblock \emph{SIAM Journal on Control and Optimization}, 30\penalty0
  (4):\penalty0 838--855, 1992.

\bibitem[Zombori et~al.(2019)Zombori, Csisz\'arik, Michalewski, Kaliszyk, and
  Urban]{zombori2019curriculum}
Zsolt Zombori, Adri\'an Csisz\'arik, Henryk Michalewski, Cezary Kaliszyk, and
  Josef Urban.
\newblock Curriculum learning and theorem proving.
\newblock In \emph{AITP}, 2019.

\end{thebibliography}

\end{document}